\documentclass[superscriptaddress,twocolumn,prl,showpacs,preprintnumbers,amsmath,amssymb,floatfix]{revtex4-1}

\usepackage{graphicx,url,braket,xcolor,xspace}
\usepackage{dcolumn}
\usepackage{bm}
\newcolumntype{.}{D{.}{.}{-1}} 
\usepackage{hyperref}
\usepackage[normalem]{ulem}

\newcommand{\be}{\begin{equation}}	
\newcommand{\ee}{\end{equation}}

\newcommand{\bbz}{$0\nu\beta\beta$\xspace}
\newcommand{\bbt}{$2\nu\beta\beta$\xspace}
\newcommand{\vlk}{$V_{{\rm low}\,k}$\xspace}
\newcommand{\Se}{$^{82}$Se\xspace}
\newcommand{\Ge}{$^{76}$Ge\xspace}
\newcommand{\Ca}{$^{48}$Ca\xspace}

\newcommand{\kev}{\, \text{keV}}

\newcommand{\X}{\hat X}

\begin{document}

\title{New determination of double-$\beta$-decay properties in $^{48}$Ca:
high-precision $Q_{\beta\beta}$-value measurement and improved nuclear matrix element calculations}

\author{A. A. Kwiatkowski} \email{aniak@triumf.ca}
\affiliation{TRIUMF, 4004 Wesbrook Mall, Vancouver, BC V6T 2A3, Canada}

\author{T. Brunner}
\altaffiliation{Present Address: Department of Physics, Stanford University, Stanford,
CA 94305, USA}
\affiliation{TRIUMF, 4004 Wesbrook Mall, Vancouver, BC V6T 2A3, Canada}

\author{J. D. Holt}
\affiliation{Institut f\"{u}r Kernphysik, Technische Universit\"{a}t Darmstadt, 64289
Darmstadt, Germany}
\affiliation{ExtreMe Matter Institute EMMI, GSI Helmholtzzentrum f\"{u}r
Schwerionenforschung GmbH, 64291 Darmstadt, Germany}

\author{A. Chaudhuri}
\affiliation{TRIUMF, 4004 Wesbrook Mall, Vancouver, BC V6T 2A3, Canada}

\author{U. Chowdhury}
\affiliation{TRIUMF, 4004 Wesbrook Mall, Vancouver, BC V6T 2A3, Canada}
\affiliation{Department of Physics and Astronomy, University of Manitoba, Winnipeg,
MB R3T 2N2, Canada}

\author{M. Eibach}
\affiliation{Institut f\"{u}r Kernchemie, Johannes Gutenberg-Universit\"{a}t, 55128 Mainz,
Germany}
\affiliation{Fakult\"{a}t f\"{u}r Physik und Astronomie, Ruprecht-Karls-Universit\"{a}t, 69120 Heidelberg, Germany}

\author{J. Engel}
\affiliation{Department of Physics and Astronomy, University of North Carolina,
Chapel Hill, NC 27599, USA}

\author{A. T. Gallant}
\affiliation{TRIUMF, 4004 Wesbrook Mall, Vancouver, BC V6T 2A3, Canada}
\affiliation{Department of Physics and Astronomy, University of British Columbia,
Vancouver, BC V6T 1Z1, Canada}

\author{A. Grossheim}
\affiliation{TRIUMF, 4004 Wesbrook Mall, Vancouver, BC V6T 2A3, Canada}

\author{M. Horoi}
\affiliation{Department of Physics, Central Michigan University, Mount Pleasant,
MI 48859, USA}

\author{A. Lennarz}
\affiliation{TRIUMF, 4004 Wesbrook Mall, Vancouver, BC V6T 2A3, Canada}
\affiliation{Institut f\"{u}r Kernphysik, Westf\"{a}lische Wilhelms-Universit\"{a}t,
48149 M\"{u}nster, Germany}

\author{T. D. Macdonald}
\affiliation{TRIUMF, 4004 Wesbrook Mall, Vancouver, BC V6T 2A3, Canada}
\affiliation{Department of Physics and Astronomy, University of British Columbia,
Vancouver, BC V6T 1Z1, Canada}

\author{M. R. Pearson}
\author{B. E. Schultz}
\author{M. C. Simon}
\affiliation{TRIUMF, 4004 Wesbrook Mall, Vancouver, BC V6T 2A3, Canada}

\author{R.A. Senkov}
\affiliation{Department of Physics, Central Michigan University, Mount Pleasant,
MI 48859, USA}

\author{V. V. Simon}
\altaffiliation{Present Address: Helmholtz-Institut Mainz, 55128 Mainz, Germany}
\affiliation{TRIUMF, 4004 Wesbrook Mall, Vancouver, BC V6T 2A3, Canada}
\affiliation{Max-Planck-Institut f\"{u}r Kernphysik, Saupfercheckweg 1, 69117 Heidelberg, Germany}
\affiliation{Fakult\"{a}t f\"{u}r Physik und Astronomie, Ruprecht-Karls-Universit\"{a}t, 69120 Heidelberg, Germany}

\author{K. Zuber}
\affiliation{Institiut f\"{u}r Kern- und Teilchenphysik, Technische Universit\"{a}t Dresden, 01069 Dresden, Germany}

\author{J. Dilling}
\affiliation{TRIUMF, 4004 Wesbrook Mall, Vancouver, BC V6T 2A3, Canada}
\affiliation{Department of Physics and Astronomy, University of British Columbia, Vancouver, BC V6T 1Z1, Canada}

\date{\today}

\begin{abstract}
We report a direct measurement of the $Q_{\beta\beta}$-value of the
neutrinoless double-$\beta$-decay candidate $^{48}$Ca at the TITAN Penning-trap
mass spectrometer, with the result that $Q_{\beta\beta}=4267.98(32)\kev$.  We
measured the masses of both the mother and daughter nuclides, and in the latter
case found a 1 $\kev$ deviation from the literature value.  In addition to the
$Q_{\beta\beta}$-value, we also present results of a new calculation of the
neutrinoless double-$\beta$-decay nuclear matrix element of $^{48}$Ca.  Using
diagrammatic many-body perturbation theory to second order to account for
physics outside the valence space, we constructed an effective shell-model
double-$\beta$-decay operator, which increased the nuclear matrix element by
about 75\% compared with that produced by the bare operator.  The new
$Q_{\beta\beta}$-value and matrix element strengthen the case for a $^{48}$Ca
double-$\beta$-decay experiment.
\end{abstract}

\pacs{32.10.Bi, 23.40.Bw, 07.75.+h, 14.60.St}


\maketitle


The discovery of neutrino oscillations represents the first evidence for new
physics beyond the Standard Model~\cite{Fuk98,Ahm02}. The oscillations
conclusively demonstrate that neutrinos have mass, that flavor eigenstates
are mixtures of mass eigenstates, and that neutrino physics is more
complicated than we had thought.  The observation of neutrinoless
double-$\beta$ (\bbz) decay, extremely rare if it exists, would at once fill
multiple gaps in our understanding of the neutrino's nature and would represent
a major breakthrough for particle physics. Since this lepton-number-violating
process can occur only if the neutrino is its own antiparticle, its discovery
would unambiguously confirm the neutrino as a Majorana particle, while a
measured lifetime would provide a value for the neutrino mass
scale~\cite{avi08}.  In order to extract that value from the \bbz-decay half-life,
however, two quantities must be accurately determined: a phase-space factor,
which depends on the $Q_{\beta\beta}$-value of the decay, and a nuclear matrix
element, which is not observable and therefore must be obtained from nuclear
structure theory.

The twelve nuclides that have been observed to undergo two-neutrino
double-$\beta$ (\bbt) decay~\cite{Barabash10,Ackerman11} are the basis for a
number of large-scale experimental \bbz-decay searches currently underway. Of
these nuclides, $^{48}$Ca possesses the largest $Q_{\beta\beta}$-value of 4.3
MeV \cite{ame12}, giving it several distinct experimental advantages.  Because
the $Q_{\beta\beta}$-value lies well above the energy of naturally occurring
background, a good signal-to-noise ratio is ensured, while the large
phase-space factor enhances the \bbz-decay rate.  The low isotopic abundance of
$^{48}$Ca, however, requires enrichment. $^{48}$Ca is currently being measured at
NEMO-III~\cite{Arn05} and studied at CANDLES~\cite{Ume06} and
CARVEL~\cite{Zdesenko05}.  The $Q_{\beta\beta}$-value provides vital input for
the simulation of signal and background, the analysis of current data, and the
design of future detectors.  In order for the uncertainty of the
$Q_{\beta\beta}$-value to be negligible in these studies, the required
precision has to be better than the intrinsic resolution of the detector.

The deep implications of massive neutrinos have led to a concentrated effort to
calculate the nuclear matrix element for \bbz-decay candidates.  Various
predominantly phenomenological many-body calculations for $^{48}$Ca currently
agree to within a factor of about three~\cite{cau08,bar12}.  That uncertainty
implies the same factor of three in an extracted neutrino mass; consequently,
improved calculations are vital.  $^{48}$Ca occupies a unique position among
\bbz-decay candidates in that its relatively low mass and doubly magic nature
make it a near ideal case for several ab-initio many-body methods developed for
medium-mass nuclei.  Many calculations of ground- and excited-state energies with
two- (NN) and three-nucleon (3N) forces in the calcium region exist and agree
with each other ~\cite{hol12,hol13a,gal12,wien13,hagen12,roth12,hergert13}, but
no attempt has yet been made to calculate the \bbz-decay matrix element in
$^{48}$Ca at the same level of sophistication.  Here, using methods first
applied to $^{76}$Ge and $^{82}$Se~\cite{eng09,lin13,holtengel}, we applied
chiral nuclear forces~\cite{chiral} and diagrammatic many-body perturbation
theory to calculate an effective shell model \bbz-decay operator for
$^{48}$Ca.  We found an increase in the nuclear matrix element of $\approx75\%$
compared to that produced by the bare operator alone, and estimated a further
increase of $\approx 8\%$ from moving beyond the closure approximation.  To
derive the decay rate, the resulting nuclear matrix element is combined with
the phase space factor, which depends on the $Q_{\beta\beta}$-value to the
fifth power.  We have determined the $Q_{\beta\beta}$-value in a direct
measurement at TRIUMF's Ion Trap for Atomic and Nuclear
science (TITAN).

\begin{figure}
  \includegraphics[width=0.8\columnwidth]{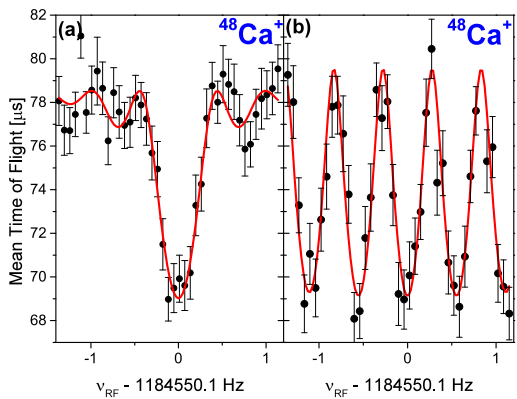}
  \caption{Typical $^{48}$Ca resonances using the TOF-ICR technique with
  $T_{RF}$ = 1.953 s (a) and Ramsey technique with an excitation scheme of
  200-1553-200 ms (b).  The solid line is an analytic fit
  \cite{koenig95,ramseyExc1} to the data.}
  \label{fig:resonances}
    \vspace{-3mm}
\end{figure}


TITAN is an ion trap system coupled to the ISAC-TRIUMF rare beam facility.  It
consists of three traps: a radiofrequency quadrupole (RFQ) beam cooler
and buncher \cite{rfq2}, an electron beam ion trap (EBIT) \cite{ebit2}, and a
measurement Penning trap (MPET) \cite{mpet}; the EBIT was not used in this experiment.  Ions were delivered from either ISAC's Off-Line
Ion Source (OLIS) \cite{olis} or TITAN's surface-ionization Ion Source (TIS).  For the
production of $^{48}$Ca$^+$, an enriched ion source was heated in the TIS whereas $^{48}$Ti$^+$ and $^{14}$N$^{18}$O$^{16}$O$^+$ ions
were produced with OLIS.  The beams from the TIS and OLIS were delivered independently.

The continuous beam from either ion source was accumulated, cooled, and
bunched in the RFQ.  A fast time-of-flight mass filter \cite{bng} placed between the RFQ and the MPET and
a dynamic capture process in the Penning trap ensured pure isobaric ion
bunches in MPET.  In addition, dipole cleaning
\cite{Bollen96} was applied to remove any remaining contaminant ions.  In a
Penning trap, the mass of an ion is measured via the determination of the cyclotron
frequency $2 \pi \nu_c = q/m \cdot B$, where $q/m$ is the charge-to-mass ratio and
$B$ the magnetic field strength.  The masses were determined using two excitation
schemes: the conventional time-of-flight ion-cyclotron-resonance (TOF-ICR)
method \cite{toficr,Bollen90}, whereby the ions were excited with a continuous RF
field for a time $T_{RF}$ (Fig. \ref{fig:resonances}a) and the Ramsey technique
\cite{ramseyExc1,ramseyExc2}, wherein the oscillatory field was applied in two
pulses separated by a waiting period (Fig. \ref{fig:resonances}b).  For this, two
200 ms RF pulses were spaced apart by 1553 ms, denoted as 200-1553-200 ms.

\begin{table}
\caption{The $Q_{\beta\beta}$ value of $^{48}$Ca and the masses of mother and daughter nuclides were found by interleaving cyclotron-frequency measurements of $^{48}$Ca$^+$, $^{48}$Ti$^+$, and N$^{18}$O$^{16}$O$^+$; the tabulated ratios are the weighted average of seven data sets.  The the total (statistical and systematic) is listed in parentheses and the statistical uncertainty in square brackets. The last column indicates the precision $\delta R/R$ achieved.}
\begin{ruledtabular}
\begin{tabular}{c l c}
 Species & Ratio & Precision \\ \hline
 $^{48}$Ca$^+$-$^{48}$Ti$^+$          & 0.999 904 448 9(46)[31] & 5 $\times$ 10$^{-9}$ \\
 $^{48}$Ca$^+$-N$^{18}$O$^{16}$O$^+$  & 1.000 930 621 6(61)[35] & 6 $\times$ 10$^{-9}$ \\
 $^{48}$Ti$^+$-N$^{18}$O$^{16}$O$^+$  & 1.001 026 276 8(47)[41] & 5 $\times$ 10$^{-9}$ \\
\end{tabular}
\end{ruledtabular}
\label{tab:ratios}
  \vspace{-3mm}
\end{table}
\begin{table*}
  \caption{A comparison of the $Q_{\beta\beta}$ and mass excesses (ME) determined
  in this work to recent values.  ISOLTRAP had determined the mass of TiO using
  the reference masses $^{85}$Rb and $^{55}$Mn as -48492.9(1.0) and
  -48492.5(1.2) keV respectively; the weighted average is listed below.  All values
  are in keV.}
  \begin{ruledtabular}
  \begin{tabular}{c . . . . .}
  & \multicolumn{1}{c}{TITAN} & \multicolumn{1}{c}{LEBIT} & \multicolumn{1}{c}{ISOLTRAP} & \multicolumn{1}{c}
  {AME 2003} & \multicolumn{1}{c}{AME 2012}  \\ \hline
  Q$_{\beta\beta}$ &   4267.98(32) &   4268.121(79)  &             &   4273.60(4.00) &   4266.98(38) \\
  ME($^{48}$Ca)    & -44224.45(27) & -44224.767(194) &             & -44214(4)       & -44224.759(120) \\
  ME($^{48}$Ti)    & -48492.70(21) &                 & -48492.3(8) & -48487.7(8)     & -48491.734(358) \\
  Ref.             & \multicolumn{1}{c}{this work} & \multicolumn{1}{c}{\cite{lebitCa48,lebitQ}} & \multicolumn{1}{c}{\cite{isoltrapTi48}}
  & \multicolumn{1}{c}{\cite{ame03}} & \multicolumn{1}{c}{\cite{ame12}} \\
  \end{tabular}
  \end{ruledtabular}
  \label{tab:masses}
  \vspace{-3mm}
\end{table*}
The $\nu_c$ measurements of $^{48}$Ca$^+$, $^{48}$Ti$^+$, and N$^{18}$OO$^+$ were interleaved; thus, the primary experimental result is the ratio of their cyclotron frequencies, listed in Tab. \ref{tab:ratios}.  A statistical uncertainty of $\delta R/R$ = 3$\cdot$10$^{-9}$ was achieved.  Systematic uncertainties were carefully evaluated.  These include simultaneous storage of multiple ions, either of the same or different species.  To determine the influence of ion-ion-interactions \cite{Bollen92}, we analyzed the data considering only events of one detected ion.  Moreover, a count-class analysis \cite{Kellerbauer03} was applied and, to be conservative, we added the difference in the ratios in quadrature to the statistical uncertainty.  In addition, nonlinear decay in the magnetic field may cause shifts in the system of 0.04(11) ppb/h \cite{li6}; as measurements with $T_{RF}$ $\approx$ 2 s were separated by approximately 1.5 hours, a 0.23 ppb correction was included.  Further off-line studies revealed frequency shifts on the level of 1.3 ppb as a result of unbalanced RF excitation stemming from instabilities in the frequency generator trigger.  As all measured ions were isobars, with identical nominal $m/q$, they followed the same nominal ion trajectory and experienced the same magnetic and electric fields.  Thus, relativistic effects and any mass-dependent effects canceled in the ratio.  We varied the excitation times (for conventional excitations $T_{RF}$ = 0.457, 1.913, 1.953 s; for Ramsey 150-653-150 and 200-1553-200 ms) for different data sets to investigate excitation-scheme dependent effects.  In addition, the time window allowed for the dynamic capture of the ion bunch in the Penning trap, was varied by -0.5 $\mu$s and +0.3 $\mu$s from the optimal value to verify the trap compensation (see e.g. \cite{mpet}).  No statistically significant differences were observed for any of these variations, and all data sets were included in the weighted average. All systematic uncertainties were added in quadrature to the statistical uncertainty and are included in Tab. \ref{tab:ratios}.

The ratios $R$ can be related to the $Q_{\beta\beta}$-value and were used to find the masses of Ca and Ti from that of N$^{18}$OO by
\begin{equation}
\label{eq:RtoQ}
Q_{\beta\beta} = (R - 1) (M_{Ti} - m_e) + R B_{Ti} - B_{Ca}
\end{equation}
\begin{equation}
M_{Ca,Ti} = R (M_{N^{18}OO} - m_e) + m_e + R B_{N^{18}OO} - B_{Ca,Ti}.
\label{eq:RtoM}
\end{equation}
where $M$ refers to the atomic mass, $m_e$ the electron mass, $B$ the electronic binding energy of the outermost electron and the subscripts identify the nuclide.  Values for $B$ were taken from \cite{nist}.  Table \ref{tab:masses} compares the values achieved in this work with values found in recent literature.

The following results could be extracted:  We determined for the first time the atomic mass of $^{48}$Ti directly using Penning trap mass spectrometry and found the mass excess to be -48492.71(21) keV; this is a 2.2$\sigma$ deviation from the Atomic Mass Evaluation (AME) 2012.  We also confirm the mass measurement of $^{48}$Ca of \cite{lebitCa48}, which deviates 10.6(4.1) keV from the previous evaluation in 2003 \cite{ame03}.  Finally, we measured the $Q_{\beta\beta}$-value, the most relevant parameter for the \bbz decay, to be 4267.98(32) keV from direct frequency ratios.  This value disagrees with the $Q_{\beta\beta}$-value as evaluated in AME 2012, which is based off the Penning-trap mass measurement of $^{48}$Ca and indirect mass measurements of $^{48}$Ti.  Prior to AME 2012, the ISOLTRAP collaboration measured $M$($^{48}$Ti) with $^{48}$TiO molecules and found a value in agreement with the TITAN value.  That is, previously the calculated $Q_{\beta\beta}$-value depended on which mass and reaction values were taken, whereas our value is directly and self-consistently determined.  More recently, the $Q_{\beta\beta}$-value was measured at the LEBIT facility \cite{lebitQ}.  Our result is in excellent agreement with theirs.  With consideration of the LEBIT and ISOLTRAP measurements, we have unambiguously determined that the shift in the $Q_{\beta\beta}$-value is due to an error in the previously accepted atomic mass value of $^{48}$Ti.


With an accurate determination of the $Q_{\beta\beta}$-value (and hence
phase-space factor), the final ingredient to connect the \bbz-decay rate with the
neutrino mass is a nuclear matrix element governing the decay.  The matrix
element is given by
\be
M_{0\nu}=M_{0\nu}^{GT}-\frac{g^2_V}{g^2_A}M^F_{0\nu}+M^T_{0\nu}
\ee
where $g_V$ and $g_A$ are the axial and vector coupling constants, and in
addition to the usual Gamow-Teller and Fermi terms, we also include the tensor
part, which has been shown to be non-negligible in \Ca~\cite{men08}.  Of the
theoretical methods used to calculate this matrix element, only the nuclear
shell model provides an exact treatment of many-body correlations, albeit
within a truncated single-particle space (valence space) above an inert
core.  Though nearly all shell-model Hamiltonians to date rely on
phenomenological adjustments to mimic correlations outside the valence space,
no modifications are made to the \bbz-decay operator. The effect of
correlations outside the valence space on the \bbz-decay nuclear matrix element
thus remains an open question.

Many-body perturbation theory (MBPT) provides a diagrammatic prescription to
account for excitations outside the valence space directly from nuclear
forces~\cite{kuo91,hjo95}.  When carried out to sufficiently high order,
diagonalization of the resulting {\it effective} valence-space Hamiltonian,
$H_{\rm eff}$, will reproduce exactly a subset of eigenvalues of the full
$A$-body problem (provided the series converges).  Despite its long history and
recent success in producing ab-initio valence-space Hamiltonians from NN and 3N
forces~\cite{hol12,hol13a}, MBPT is only now being extended to calculate
effective two-body operators~\cite{eng09,lin13,holtengel}.  We applied this
formalism to construct an effective valence-space \bbz-decay operator for \Ca.

We took as our valence space the standard $pf$ shell, consisting of $f_{7/2}$,
$p_{3/2}$, $p_{1/2}$, $f_{5/2}$ orbitals above a $^{40}$Ca core.  We first
constructed the \emph{$\X$}-box, an object which includes all ``unfolded''
diagrams containing the \bbz-decay transition operator~\cite{holtengel}.  At
lowest order, $\X$ is the bare \bbz-decay operator, and in the current work, we
truncated $\X$ at second-order in the nuclear interaction.  To obtain the final
effective \bbz-decay operator, we included once-folded $\X$-box diagrams and
state norms as in Ref.\ \cite{holtengel}.  The interaction in these diagrams
was the NN force derived from chiral effective field theory(EFT) at order
N$^3$LO~\cite{n3lo} and evolved to low momentum (yielding the potential \vlk)
via renormalization group methods~\cite{vlowk}.  To obtain the nuclear matrix
element itself, we combined our effective operator with wave functions
calculated from the GXPF1A interaction~\cite{GXPF}.

\begin{table}
\centering
\begin{tabular*}{0.48\textwidth}{@{\extracolsep{\fill}}lcccc}\hline\hline
& $M^{GT}_{0\nu}$ & $-\frac{g_V^2}{g_A^2}M^{F}_{0\nu}$ & $M^{T}_{0\nu}$&Sum
\\[1mm]
\hline
Bare matrix element $M_{0\nu}$& 0.675  & 0.130 & $-0.072$ & 0.733 \\[.5mm]
First-order $\X$-box, no 3p-1h & 1.340& 0.225 & $-0.064$&1.501\\[.5mm]
Full first-order $\X$-box  & 0.616 & 0.125 & $-0.123$ & 0.619 \\[.5mm]
Full second-order $\X$-box& 1.822 & 0.233 & $-0.063$& 1.992 \\[.5mm]
\hline
Final matrix element & 1.211 & 0.160 & $-0.070$ & 1.301 \\[.5mm]

\hline\hline
\end{tabular*}
\caption{The \bbz-decay matrix elements $M_{0\nu}$ for \Ca at various
approximations in our many-body framework.}
\label{tab:nme}
\vspace{-3mm}
\end{table}

Our results for the \Ca nuclear matrix element appear in Tab.~\ref{tab:nme},
where we list the contributions to the different parts of the operator at
various orders in \vlk. We see the same trends as in \Se and \Ge, namely, at
first-order, ladder effects increase the total matrix element by a factor of
two, followed by a significant reduction from core-polarization diagrams.
Here, however, the effects of second-order diagrams ($\approx120$ in all) and
folding are larger, yielding a final value $\approx75\%$ larger than that
obtained from the bare \bbz-decay operator alone. (The increase in \Ge and \Se
was less than half as much.)  We also found that the bare matrix element
increased by 8\% when we avoided the closure approximation.  Although we cannot
avoid closure for our effective operator, its matrix element would likely
increase by a similar amount.

Though these calculations represent significant progress towards a fully
ab-initio calculation and offer our best estimate for the nuclear matrix
element with \Ca, the large second-order contributions to $\X$ mean that
higher-order contributions could also be significant.  Pushing to higher order
will be difficult, but we plan other improvements: replacing the
phenomenological wavefunctions by those obtained from an ab-initio $H_{\rm
eff}$, including the effects of two-body weak currents in the bare
operator~\cite{men11}, investigating the size of induced three-body operators~\cite{shu11},
and including 3N forces in intermediate-state $\X$-box excitations.

In conclusion, we provided two improved quantities for $^{48}$Ca \bbz decay that together are required to extract the effective Majorana neutrino mass from the decay rate.  The $Q_{\beta\beta}$-value is now precisely
determined in a self-consistent way and confirm a large deviation from separate
determinations.  The discrepancy with the accepted $^{48}$Ti mass value, uncovered in the recent LEBIT $Q_{\beta\beta}$-value measurement, has been resolved by our mass measurement, revealing a shift of $\approx$1 $\kev$. In addition, we obtained the nuclear matrix element by including the
effects of levels outside the valence space in a shell model calculation.
These efforts make a \bbz experiment in $^{48}$Ca more attractive.


\begin{acknowledgments}
The authors are indebted to K. Jayamanna for operation of OLIS.  We thank D.
Frekers and J.\ Men\'endez for many fruitful discussions.  We thank M. Good for
his assistance in upgrades to the TIS.  This work has been supported by the
Natural Sciences and Engineering Research Council (NSERC) of Canada and the
National Research Council (NRC) of Canada through TRIUMF. J.E. acknowledges
support from the U.S. Department of Energy under Contract No.\
DE-FG02-97ER41019.  A.T.G. acknowledges support from the NSERC CGS-D, A.L. from
the Deutsche Forschungsgesellschaft (DFG) under grant number no. FR601/3-1,
T.D.M from NSERC CGS-M, and V.V.S. from the Studienstiftung des deutschen
Volkes.  This work was supported by BMBF under Contract No.~06DA70471 and the
Helmholtz Association through the Helmholtz Alliance program, Contract
No.~HA216/EMMI ``Extremes of Density and Temperature: Cosmic Matter in the
Laboratory''.
\end{acknowledgments}

\bibliographystyle{apsrev}
\bibliography{48Ca_vbb_submit}

\end{document}